\documentclass[prl,showpacs,amsfonts,amssymb,amsmath,tightenlines,nofootinbib,showkeys,twocolumn]{revtex4}

\usepackage{amssymb, amsmath, amsfonts, latexsym, verbatim, mathrsfs}

\usepackage{graphicx}

\newtheorem{theorem}{Theorem}
\newtheorem{itlemma}{Lemma}[section]
\newtheorem{itproposition}[itlemma]{Proposition}
\newtheorem{itcorollary}[itlemma]{Corollary}
\newtheorem{itremark}[itlemma]{Remark}
\newtheorem{itremarks}[itlemma]{Remarks}
\newtheorem{itdefinition}{Definition}
\newtheorem{itexample}[itlemma]{Example}

\newenvironment{definition}{\begin{itdefinition}\rm}{\end{itdefinition}}

\begin{document}

\title{Environment induced incoherent controllability}
\thanks{Work supported by NSF Career Grant, ECS0237925}

\author{Raffaele Romano}
\email{rromano@iastate.edu}
\author{Domenico D'Alessandro}
\email{daless@iastate.edu} \affiliation{Department of Mathematics,
Iowa State University, Ames IA 50011, USA}


\begin{abstract}

\noindent We prove that the environment induced entanglement between
two non interacting, two-dimensional quantum systems $S$ and $P$ can
be used to control the dynamics of $S$ by means of the initial state
of $P$. Using a simple, exactly solvable model, we show that both
accessibility and controllability of $S$ can be achieved under
suitable conditions on the interaction of $S$ and $P$ with the
environment.

\end{abstract}



\maketitle


\section{Introduction}
\label{sec1}

Control theoretical methods and concepts are playing an important
role in the development of modern quantum mechanics and in
particular of quantum information theory~\cite{niel}. Control theory
ideas are used both in the analysis of quantum dynamics and
generation of entanglement~\cite{turc,cabr,sore,sock,beig} and in
the development of algorithms for the control of quantum
systems~\cite{pala,tesc,garc,rang}. In this context, a fundamental
question is to what extent it is possible to influence the dynamics
of  a quantum system by an external action. This {\it
controllability} question will be the object of the present letter.

It is assumed that the dynamics of a quantum system $S$ depends on a
number of  parameters $u$ (the {\it controls}) that can be
externally modified, i.e.
\begin{equation}\label{dyn}
    \rho_S (t, u) = \gamma (t, u) [\rho_S(0)]
\end{equation}
\noindent for some linear map $\gamma=\gamma(t,u)$, where $\rho_S$
is the statistical operator associated to $S$.

In the controllability analysis of a given system, one wishes  to
study the set of the possible transitions that can be induced in the
system by choosing the controls $u$. To formalize this question,
one  introduces  the {\it reachable set from $\rho_S$ at time t} as
\begin{equation}\label{reach}
    {\cal R}(\rho_S, t) = \{\rho_S (t, u) \vert \rho_S (0) =
\rho_S, u \in {\cal
    U}\},
\end{equation}
where ${\cal U}$ is the set of admitted controls. The {\it reachable
set from $\rho_S$} is given by
\begin{equation}\label{reach2}
    {\cal R}(\rho_S) = \lim_{T \rightarrow + \infty} {\cal R}_T(\rho_S),
\end{equation}
where
\begin{equation}\label{reach3}
    {\cal R}_T (\rho_S) = \bigcup_{0 \leqslant t \leqslant T} {\cal R}(\rho_S, t)
\end{equation}
is the reachable set from $\rho_S$ until $T$. In general ${\cal
R}(\rho_S) \subseteq {\cal P}_S$, where ${\cal P}_S$ is the convex
set of all density matrices associated to $S$. The main
controllability properties are defined in connection to these sets
as follows.
\begin{definition}\label{def1}
The system $S$ is said {\it accessible} if and only if ${\cal
R}_T(\rho_S)$ contains non empty open sets of ${\cal P}_S$ for all
$T$ and for all $\rho_S \in {\cal P}_S$.
\end{definition}
From a physical point of view, this means that it is possible to
move every initial system $\rho_S$ in arbitrary directions in ${\cal
P}_S$ by suitably choosing the control parameters $u$.
\begin{definition}\label{def2}
The system $S$ is said {\it controllable} if and only if ${\cal
R}(\rho_S) = {\cal P}_S$ for all initial state $\rho_S \in {\cal
P}_S$.
\end{definition}
Consequently, for a controllable system $S$ every transition between
two arbitrary states in ${\cal P}_S$ is allowed.

Controllability has been investigated in depth for quantum systems
when the controls $u$ appear as parameters in the  Hamiltonian of
the system. This study has concerned  both for unitary and
dissipative evolutions and has led to several algebraic criteria to
test controllability (see e.g.~\cite{albe,alta,tarn,rama}). Methods
of control which use tunable parameters in the Hamiltonian of the
systems are referred to as {\it coherent control} methods. Motivated
by several experimental scenarios, control techniques have recently
been investigated where the control variable affect an auxiliary
system which is then made interact with the quantum system to obtain
control~\cite{vile,mand,roma}. More precisely, the system $S$ is
allowed to interact with a second system $P$, called {\it probe},
and initially they are in an uncorrelated state $\rho_S \otimes
\rho_P$. It is assumed that it is possible to modify the initial
state of $P$ before the interaction, therefore in this case the
controls enter the dynamics of $S$ through $\rho_P = \rho_P (u)$.
This control method is referred to as {\it incoherent control} since
it does not rely on modifications of the Hamiltonian of the system
$S$.

Controllability and accessibility of $S$  in the  incoherent control
setting have been investigated under the assumption that the
composite system $T = S + P$ is closed. In this case the dynamics
(\ref{dyn}) is given by \begin{equation}\label{trp}
    \rho_S (t, u) = Tr_P \Bigl(X(t)
\rho_S \otimes \rho_P (u) X^{\dagger} (t) \Bigr),
\end{equation}
where $Tr_P$ is the partial trace over the degrees of freedom of
$P$, $X(t) = e^{-i H_T t}$ is the unitary propagator and $H_T = H_S
+ H_P + H_{SP}$ is the Hamiltonian of $T$. The coupling between $S$
and $P$ is given by the interaction Hamiltonian $H_{SP}$. Necessary
and sufficient condition for controllability and accessibility have
been derived in the case of two-dimensional $S$ and $P$~\cite{roma},
under the hypothesis that it is possible to obtain all the pure
states of $\rho_P$ by an arbitrary choice of the control.

\begin{theorem}\label{Thm1}~\cite{roma} The system $S$ evolving under
(\ref{trp}) is (incoherent) controllable if and only if there is a
time $t$ at which the unitary evolution of the composite system $X
(t)$ is locally equivalent to the SWAP operator.
\end{theorem}

Algebraic conditions of incoherent controllability equivalent to the
ones expressed in Theorem~\ref{Thm1} can be given by considering the
Cartan decomposition~\cite{helg} of $X (t)$,
\begin{equation}\label{cart}
    X (t) = L_1 (t) e^{a t} L_2 (t),
\end{equation}
where $L_1 (t), L_2 (t)$ are local transformations, $a = c_x
\sigma_x^S \otimes \sigma_x^P + c_z \sigma_y^S \otimes \sigma_y^P +
c_z \sigma_z^S \otimes \sigma_z^P$ is an element of the Cartan
subalgebra of $\mathfrak{su}(4)$, $c_i$ are real coefficients and
$\sigma_i^{S, P}$ are the Pauli matrices in $S$, $P$ respectively.

Conditions for accessibility of $S$ can  be  expressed in terms the
coefficients appearing in the Cartan decomposition.

\begin{theorem}\label{Thm2}~\cite{roma}
The system $S$ evolving under (\ref{trp}) is accessible if and only
if $c_i \ne 0$ for all $i = x, y, z$.
\end{theorem}

The assumption that $T$ is a closed system is valid only in first
approximation. In general, there will be an external environment $E$
interacting with $T$ and thus affecting the controllability
properties of $S$. Since in general there is no control on $E$,
intuition suggests that the interaction between $E$ and $T$ is
always a negative factor for the controllability properties of $S$,
as it leads to a dissipative evolution for $T$. The main goal  of
this letter is to prove that this is not always true and that the
interaction with $E$ can have a positive impact for the incoherent
controllability of $S$ by $P$. In the following we shall consider a
common model for the environment given by a large number of
decoupled harmonic oscillators and show that for appropriate forms
of the bath-system interaction we can have accessibility and
controllability of a system which would otherwise be not
controllable and not accessible as a closed system. Our research is
related to the investigation in~\cite{brau1} where it was shown that
the interaction with a common environment can generate entanglement
for a couple of systems plunged in it. The case treated here is in
essence the opposite of the one treated in~\cite{roma}. In that
paper system and probe were assumed interacting and no environment
was present. In the case treated here, system and probe are assumed
not directly interacting and their interaction is totally due to the
presence of the environment.


\section{A model of two systems interacting through the
environment} \label{sec2}

We consider the model of the environment described in~\cite{brau2}.
$E$ given by a set of $N$ decoupled harmonic oscillators with
Hamiltonian
\begin{equation}\label{he}
    H_E = \sum_{i = 1}^N \hbar \omega_i \Bigl( b_i^{\dagger} b_i + \frac{1}{2} \Bigr)
\end{equation}
where $b_i^{\dagger}$, $b_i$ are the creation and annihilation
operators associated to the $i-$th oscillator and $\omega_i$ its
angular frequency. This is the bosonic bath model as $N
\rightarrow \infty$ and the considerations on  controllability
that will follow do not depend on  $N$. We assume  $H_T = 0$, that
is the composite system of system and probe, $T = S + P$, has no a
free evolution. We assume a linear coupling between $E$ and $T$
depending on the positions of the oscillators,
\begin{equation}\label{het}
    H_{ET} = \sum_{i = 1}^N A_T \otimes g_i (b_i + b_i^{\dagger}),
\end{equation}
where $g_i$ is the coupling constant of the $i-$th oscillator and
$A_T$ an arbitrary hermitian operator in the Hilbert space of $T$.
The evolution of a state of $S$ is given by
\begin{equation}\label{trp1}
    \rho_S (t, u) = Tr_P Tr_E \bigl( X (t) \rho_S
    \otimes \rho_P (u) \otimes \rho_E X^{\dagger} (t) \bigr)
\end{equation}
where $X(t) = e^{-i(H_E + H_{ET})t}$ and $S$, $P$ and $E$ are all
initially decoupled. The environment is in the thermal state
$\rho_E$. $A_T$ is a constant of motion since $[A_T, H_E + H_{ET}] =
0$, therefore it is possible to find the exact analytical expression
of the dynamics. It is convenient to introduce the eigenvalues and
eigenvectors of $A_T$, $A_T \vert \alpha_i \rangle = \alpha_i \vert
\alpha_i \rangle$ for $i = 1, \ldots , 4$, therefore (\ref{trp1})
becomes
\begin{equation}\label{trp2}
    \rho_S (t, u) = \sum_{i, j = 1}^4 Tr_P \vert \alpha_i \rangle
    \langle \alpha_j \vert \, \bigr( \rho_S \otimes \rho_P (u)
    \bigl)_{ij} \gamma_{ij} (t)
\end{equation}
where we introduced the functions
\begin{equation}\label{gam}
    \gamma_{ij} (t) = e^{-(\alpha_i - \alpha_j)^2
    f(t) + i (\alpha_i^2 - \alpha_j^2) \varphi (t)}
\end{equation}
and
\begin{eqnarray}\label{ffi}
    f(t) &=& \sum_{i = 1}^N \Bigl( \frac{g_i}{\hbar \omega_i} \Bigr)^2
    (1 + 2 \bar{n}_i) (1 - \cos{\omega_i t}), \nonumber \\
    \varphi (t) &=& \sum_{i = 1}^N \Bigl( \frac{g_i}{\hbar \omega_i}
    \Bigr)^2 (\omega_i t - \sin{\omega_i t}),
\end{eqnarray}
where $\bar{n}_i$ is the average thermal occupation number for the
$i-$th oscillator~\cite{brau2}.

To compute the partial trace in (\ref{trp2}) we need to make some
assumptions on the eigenvectors of $A_T$. In the study of the
incoherent controllability for this system, we find convenient to
explore two opposite cases: either all the eigenvectors are
factorized states in the Hilbert space of $S + P$, or they are
maximally entangled states. By exploring these two extreme cases we
will find examples of evolutions that are not accessible, accessible
but not controllable or controllable. This last case will prove our
claim that the environment induces incoherent controllability.


\section{Controllability and accessibility properties} \label{sec3}

We first consider the case where the eigenvectors of $A_T$, $\vert
\alpha_i \rangle$, are factorized states, i.e. $\vert \alpha_i
\rangle = \vert \alpha_k^S \rangle \otimes \vert \alpha_l^P \rangle$
with $i = (k, l)$, $i = 1, \ldots , 4$ and $k$, $l = 1, 2$, and the
sets $\{ \vert \alpha^S_1 \rangle, \vert \alpha^S_2 \rangle\}$, $\{
\vert \alpha^P_1 \rangle, \vert \alpha^P_2 \rangle\}$ are
orthonormal bases in the Hilbert spaces of $S$ and $P$,
respectively. In this case
\begin{equation}\label{trpa}
    Tr_P \vert \alpha_i \rangle \langle \alpha_j \vert = \delta_{ln} \vert
    \alpha_k^S \rangle \langle \alpha_m^S \vert
\end{equation}
and moreover
\begin{equation}\label{coeff}
    \bigr( \rho_S \otimes \rho_P (u) \bigl)_{ij} = (\rho_S)_{km}
    \bigl( \rho_P (u) \bigr)_{ln}
\end{equation}
where $i = (k, l)$ and $j = (m, n)$. Thus equation (\ref{trp2})
becomes
\begin{eqnarray}\label{trp3}
    \rho_S (t, u) &=& \sum_{k, m = 1}^2 \Bigl[ (\rho_S)_{km} \vert \alpha^S_k
    \rangle \langle \alpha^S_m \vert \cdot \nonumber \\
    && \cdot \sum_{n = 1}^2 \bigr(\rho_P (u)\bigr)_{nn} \gamma_{(k,n)(m,n)}(t) \Bigr]
\end{eqnarray}
and initial states $\rho_S$ that are diagonal in the considered
basis do not evolve. Examples of evolutions displaying this behavior
are determined by interaction terms of the form $A_T = A_S + A_P$ or
$A_T = A_S \otimes A_P$, where $A_S$ and $A_P$ are hermitian
operators acting on the Hilbert spaces of $S$ and $P$. It follows
that in these cases $S$ is neither accessible nor controllable,
therefore a necessary condition for accessibility and
controllability is that at least one eigenvector of $A_T$ is an
entangled state in $S + P$.

\noindent We assume now that all the eigenvectors are maximally
entangled states, i.e. Bell states
\begin{eqnarray}\label{bell}
    \vert \alpha_{1, 2} \rangle = \frac{1}{\sqrt{2}} (\vert \alpha_1^S \rangle \otimes \vert
    \alpha_1^P \rangle \pm \vert \alpha_2^S \rangle \otimes \vert \alpha_2^P
    \rangle) \nonumber \\
    \vert \alpha_{3, 4} \rangle = \frac{1}{\sqrt{2}} (\vert \alpha_1^S
    \rangle \otimes \vert
    \alpha_2^P \rangle \pm \vert \alpha_2^S \rangle \otimes \vert \alpha_1^P
    \rangle)
\end{eqnarray}
in suitable bases $\{ \vert \alpha_1^S \rangle, \vert \alpha_2^S
\rangle \}$ and $\{ \vert \alpha_1^P \rangle, \vert \alpha_2^P
\rangle \}$. It is convenient to use a coherence vector
representation for the states in $S$ and $P$, that is
\begin{equation}\label{cove}
    \rho_S = \frac{1}{2} ({\mathbb I} + \vec{s} \cdot
    \vec{\sigma}^S), \quad \rho_P = \frac{1}{2} ({\mathbb I} + \vec{p} \cdot
    \vec{\sigma}^P)
\end{equation}
where $\vec{s}$, $\vec{p}$\, are real vectors in the Bloch spheres
of $S$ and $P$ and $\vec{\sigma}^{S, P}$ are the vectors of the
Pauli matrices in $S$ and $P$. In this representation the dynamics
(\ref{trp2}) takes the form
\begin{equation}\label{dync}
    \vec{s} (t, u) = A(t, \vec{s}_0) \vec{p} (u) + \vec{a} (t, \vec{s}_0)
\end{equation}
where $A(t, \vec{s}_0)$ is the matrix
\begin{equation}\label{amat}
    \frac{1}{2} \, {\rm \mathbb Im} \left(%
    \begin{array}{ccc}
    i \gamma_{13 - 24} (t) &
    s_z \gamma_{13 - 24} (t) &
    s_y \gamma_{13 + 24} (t) \\
    s_z \gamma_{14 - 23} (t) &
    i\gamma_{23 - 14} (t) &
    - s_x \gamma_{23 + 14} (t) \\
    - s_y \gamma_{12 + 34} (t) &
    s_x \gamma_{34 - 12} (t) &
    i \gamma_{12 - 34} (t) \\
    \end{array}%
\right)
\end{equation}
and
\begin{equation}\label{ave}
    \vec{a} (t, \vec{s}_0) = \frac{1}{2} \, {\rm \mathbb Re} \left(%
\begin{array}{c}
  s_x \gamma_{13 + 24} (t) \\
  s_y \gamma_{23 + 14} (t) \\
  s_z \gamma_{12 + 34} (t) \\
\end{array}%
\right).
\end{equation}
Here, we have introduced the convenient notation
\begin{equation}\label{newg}
    \gamma_{ij \pm kl} (t) = \gamma_{ij} (t) \pm \gamma_{kl} (t)
\end{equation}
where the $\gamma_{ij}$ have been defined in (\ref{gam}), and
$\vec{s}_0 = (s_x, s_y, s_z)$ represents the initial state $\rho_S$.

Assuming that the initial state $\rho_P (u)$ can be an arbitrary
state in the Bloch sphere of $P$, it follows that in the coherence
vector formalism ${\cal R}(\rho_S, t)$ is represented by an
ellipsoid contained in the Bloch sphere of $S$, centered in
$\vec{a}(t, \vec{s}_0)$, with the semi axes given by the singular
values of $A(t, \vec{s}_0)$. A sufficient condition for
accessibility which is generically satisfied can be given in terms
of the eigenvalues of $A_T$, $\alpha_i$, $i=1,\ldots,4$. In
particular, the system is accessible if
\begin{eqnarray}\label{conda}
  (\alpha_2 - \alpha_4)^2 &\ne& (\alpha_1 - \alpha_3)^2 \nonumber \\
  (\alpha_1 - \alpha_4)^2 &\ne& (\alpha_2 - \alpha_3)^2 \nonumber \\
  (\alpha_3 - \alpha_4)^2 &\ne& (\alpha_1 - \alpha_2)^2.
\end{eqnarray}
To see that this is a sufficient condition for accessibility, one
calculates the $6-$th derivative with respect to time of the
determinant of the matrix $A(t, \vec s_0)$ in (\ref{amat}) for $t=0$
(all the lower order derivatives are zero at that point). If
condition (\ref{conda}) is verified, then this derivative is
different from zero. If the system were not accessible then the
matrix $A(t,\vec s_0)$ should have a singular value equal to zero
for every $t$ in an arbitrarily small interval $[0,\epsilon)$.
Therefore $det A(t,\vec s_0) \equiv 0$ for $t \in [0, \epsilon)$ and
its derivatives should all vanish at $t = 0$.

To find conditions for controllability is more complicate. In fact,
the evolution of the reachable set can be quite involved and the
condition ${\cal R}(\rho_S) = {\cal P}_S$ could  depend on the
parameters of the model in non trivial ways. Our goal here however
is to show that it is possible to use the environment to have
complete control over the state of the system. Therefore we will show
that incoherently controllable evolutions are possible by exhibiting
an explicit example. The simplest cases arise when ${\cal R}(\rho_S,
\hat{t}) = {\cal P}_S$ at some time $\hat{t}$. This can be obtained
by choosing $\alpha_1 = \alpha_2 = \alpha_3 = 0$ and $\alpha_4 \ne
0$. Since relations (\ref{conda}) are satisfied, we have
accessibility for all $\alpha_4 \ne 0$. Moreover, equations
(\ref{amat}) and (\ref{ave}) simplify to
\begin{equation}\label{amat2}
    A (t, \vec{s}_0) = \frac{1}{2} \left(
        \begin{array}{ccc}
        1 - \gamma_r(t) & - s_z \gamma_i(t) & s_y \gamma_i(t) \\
        s_z \gamma_i(t) & 1 - \gamma_r(t) & - s_x \gamma_i(t) \\
        - s_y \gamma_i(t) & s_x \gamma_i(t) & 1 - \gamma_r(t) \\
        \end{array}
    \right)
\end{equation}
and
\begin{equation}\label{ave2}
    \vec{a} (t, \vec{s}_0) = \frac{1}{2} \bigl( 1 + \gamma_r(t) \bigr) \left(
        \begin{array}{c}
        s_x \\
        s_y \\
        s_z \\
        \end{array}
    \right)
\end{equation}
where
\begin{eqnarray}\label{dega}
    \gamma_r(t) = e^{- \alpha_4^2 f(t)} \cos{\bigl( \alpha_4^2 \varphi
    (t) \bigr)}, \nonumber \\
    \gamma_i(t) = e^{- \alpha_4^2 f(t)} \sin{\bigl( \alpha_4^2 \varphi
    (t) \bigr)}.
\end{eqnarray}
A sufficient condition for controllability is that $A (\hat{t},
\vec{s}_0) = {\mathbb I}$ and $a (\hat{t}, \vec{s}_0) = \vec{0}$ at
some time $\hat{t}$. Therefore $\gamma_r(\hat{t}) = -1$ and
$\gamma_i(\hat{t}) = 0$, that is
\begin{equation}\label{condom}
    \left\{
      \begin{array}{ll}
        \alpha_4^2 f(\hat{t}) = 0 & \hbox{} \\
        \alpha_4^2 \varphi (\hat{t}) = (2 k_1 + 1) \pi, & \hbox{$k_1 \in {\mathbb Z}$}
      \end{array}
    \right.
\end{equation}
The first condition in (\ref{condom}) is satisfied if and only if
$\cos{\omega_i \hat{t}} = 1$, that is $\omega_i \hat{t} = 2 {k_{2i}}
\pi$ for all $i = 1, \ldots N$, with $k_{2i} \in {\mathbb Z}$. This
condition can certainly be satisfied if the environment consists of
a field in a cavity, for appropriate values of $k_{2i}$.  Finally,
using the second equation with $\sin{\omega_i \hat{t}} = 0$ for all
$i = 1, \ldots N$, we find a condition on the eigenvalue $\alpha_4$:
\begin{equation}\label{a4}
    \frac{1}{\alpha_4^2} = \frac{2}{2 k_1 + 1}
\sum_{i = 1}^N k_{2i}\Bigl( \frac{g_i}{\hbar \omega_i}
    \Bigr)^2,
\end{equation}
with arbitrary $k_1 \in {\mathbb Z}$. Therefore, controllability can
be achieved for an appropriate combination of the parameters
defining the dynamics of the bath (the frequencies $\omega_i$) and
the parameters defining the interaction (the $\alpha_j$'s,
$j=1,\ldots,4$).


Condition~(\ref{a4}) is a rather strict request on the coefficient
$\alpha_4$. However, if it is possible to change some parameters in
the bath dynamics (e.g. the intensity of the electromagnetic field
in a cavity) they could be tuned in order to realize an incoherently
controllable system.

The crucial point to obtain controllability is that the interaction
of the environment with the system $T$ must have at least one
entangled eigenvector. A physical example is given by two identical
quantum dots, localized in different positions $q_S$ and $q_P$, in
an electromagnetic cavity, with a non-dipole interaction with the
electromagnetic field. The position degrees of freedom, not involved
in $A_T$, can be used to distinguish the probe from the system and
then to perform the incoherent control protocol.


\section{Conclusions}
\label{sec4}

We have described a model of incoherent control of a system $S$ by
means of a probe $P$, in the presence of a common environment $E$.
We have assumed that $S$ and $P$ do not evolve in absence of the
environment, so their dynamics is only due to the interaction with
$E$. In this framework, we have proved that the induced correlations
between $S$ and $P$ are, in some cases, rich enough to allow {\it
total} control of $S$ through $P$. These results complement recent
research on the creation of entanglement and suggests that further
investigations of the control of a quantum system through its
correlations with the environment will prove fruitful.

In this model, a necessary condition for accessibility and
controllability is that the interaction of $E$ with $T = S + P$ is
not a superposition of separate interactions. Otherwise, even if $S$
and $P$ become entangled, it is not possible to achieve
accessibility or controllability of $S$. It is still possible to
drive $S$ using $P$, but this is a limited ability.

The dynamics considered in this work is non-Markovian, but this does
not seem to be a fundamental assumption: in fact, it has been proved
in~\cite{bena} that also Markovian dynamics can entangle initially
uncorrelated systems.



\begin{thebibliography}{99}


\bibitem{niel} M.A. Nielsen and I.L. Chuang, {\it Quantum Computation and Quantum
Information}, Cambridge, 2000

\bibitem{turc} Q. A. Turchette {\it et al.}, Phys. Rev. Lett. 81, 3631 (1998)

\bibitem{cabr} C. Cabrillo, J.I. Cirac, P. Garc{\' i}a-Fern{\' a}ndez and P. Zoller, Phys.
Rev. A 59, 1025 (1999)

\bibitem{sore} A. S{\o}rensen and K. M{\o}lmer, Phys. Rev. A 62, 022311 (2000)

\bibitem{sock} C.A. Sackett {\it et al.}, Nature 404, 256 (2000)

\bibitem{beig} A. Beige {\it et al.}, J. Mod. Opt. 47, 2583 (2000)

\bibitem{pala} J.P. Palao and R. Kosloff, Phys. Rev. Lett. 89, 188301 (2002)

\bibitem{tesc} C.M. Tesch and R. de Vivie-Riedle, Phys. Rev. Lett. 89, 157901 (2002)

\bibitem{garc} J.J. Garc{\' i}a-Ripoll, P. Zoller and J.I. Cirac, Phys. Rev. Lett.
91, 157901 (2003)

\bibitem{rang} C. Rangan, A.M. Bloch, C. Monroe and P.H. Bucksbaum, Physical Review
Letters 92, 113004 (2004)

\bibitem{albe} F. Albertini and D. D'Alessandro, IEEE Transactions on Automatic
Control 48, 1399 (2003)

\bibitem{alta} C. Altafini, J. Math. Phys. 44, 2357 (2003)

\bibitem{tarn} G. M. Huang, T. J. Tarn and J. W. Clark,
J. Math. Phys. 24, 2608 (1983)

\bibitem{rama} V. Ramakrishna, M. V. Salapaka, M. Dahleh, H. Rabitz and
A. Peirce, Phys. Rev. A 51, 960 (1995)

\bibitem{vile} R. Vilela Mendes and V.I. Manko, Phys. Rev. A 67,
053404 (2003)

\bibitem{mand} A. Mandilara and J. W. Clark, Phys. Rev. A 71, 013406
(2005)

\bibitem{roma} R. Romano and D. D'Alessandro, quant-ph/0510020

\bibitem{helg} S. Helgason, {\it Differential Geometry, Lie Groups
and Symmetric Spaces}, Academic Press, 1978

\bibitem{brau1} D. Braun, Phys. Rev. Lett. 89, 277901 (2002)

\bibitem{brau2} D. Braun, F. Haake and W. T. Strunz, Phys. Rev.
Lett. 86, 2913 (2001)

\bibitem{bena} F. Benatti, R. Floreanini and M. Piani, Phys. Rev.
Lett. 91, 070402 (2003)



\end{thebibliography}
\end{document}